# Ultra-broadband Terahertz Perfect Absorber Based on Multi-frequency Destructive Interference and Grating Diffraction

Cheng Shi, XiaoFei Zang*, XueBin Ji, Lin Chen, Bin Cai and YiMing Zhu*

*Shanghai Key Lab of Modern Optical System and Engineering Research Center of Optical Instrument and System, Ministry of Education, University of Shanghai for Science and Technology, Shanghai 200093, China*

**Abstract:** High absorption in a wide frequency band has attracted considerable interest since their potential applications in frequency spectrum imaging systems and anti-radar cloaking. In this paper, a polarization-independent, ultra-broadband and omnidirectional terahertz absorber is proposed, fabricated and evaluated. It is experimentally demonstrated that an over 95% absorption can be obtained in the frequency range of 0.75~2.41 THz. Attributing to the multi-frequency destructive interference between the layers and the impedance-matching condition of the grating, five successive absorption peaks at 0.88 THz, 1.20 THz, 1.53 THz, 1.96 THz and 2.23 THz merged into a ultra-broadband absorption spectrum.

# 1. INTRODUCTION

The concept of trapping energy in a thin device has been one of the research hotspots for decades. To date, they have found significant applications in diverse areas including sensor detection,[1] power harvesting,[2] imaging spectrum system[3] and so on. With the advent of the metamaterial, nearly unity power can be captured at a certain frequency. The metamaterial absorber (MPA) was firstly reported in 2008,[4] which consists of a three-layer system that includes a subwavelength frequency selective surface (FSS) layer, a dielectric spacer layer, and a metal substrate. The reflective waves from the FSS/spacer interface and the metal plate can cause the destructive interference effect which will lead to the unity absorption at the resonance frequency.[5,6] However, a single electric ring resonator (ERR) structure in the FSS lead to only one resonance frequency.[7] Evolved from this characteristic, more FSSs with slightly different ERRs were designed in order to create a broadband absorption spectrum.[8,9] This sort of multi-FSS layers structure will make the several resonant frequencies closely positioned and formed a broader absorption spectrum. Nevertheless, such broadband absorbers are too complicated to be massively produced by the semiconductor technology, which makes the fabrication of such structure are quite challenging, especially for micro- and nanometer waves.[10] Recently, heavily-doped silicon grating terahertz absorber[11,12] and planar photonic-crystal terahertz absorber[13] are reported to achieve a high absorption for a broad bandwidth. Furthermore, we firstly found that the FSS layer in the conventional MPA structure can be substituted by the grating structure, which can not only still form the destructive interference of the multiple reflections but also obstruct the constructive interference of the reflective waves if designed appropriately. Successive absorption peaks caused by the destructive interference and the impedance-matching condition of the grating will be thus integrated into an ultra-broadband absorption spectrum. The design will substantially simplify the fabrication as well as enlarge the bandwidth of absorption spectrum. The absorber experimentally obtained an over 90% absorption ranges from 0.65 THz to 2.45 THz, whose relative absorption bandwidth (RAB) reaches up to 120%, broader than any other absorber in this frequency range to our knowledge.[14]

## 2. PRINCIPLE

The absorber is made up of three layers, which is similar to the construction of the single-resonance MPA. We evolve it through replacing the metal ERR and the polyimide spacer by the heavily-doped silicon grating and spacer (Fig.1 (a)). The geometric parameter is given as $w = 22$ μm, $l = 73$ μm, $p = 105$ μm, $t_1 = 51$ μm, $t_2 = 74$ μm, $t_3 = 200$ nm. As the thickness of the metal plate is much larger than the typical skin depth in the terahertz regime, the reflection is the only factor limiting absorption. To analyze the reflection spectrum of the device in detail, the interference model is established as illustrated in Fig. 2(a), the reflection waves of different orders can be calculated as follow:

$$\begin{aligned}
I_0 &= I_0 e^{iwt} \\
r_0 &= r_{11} I_0 e^{i(wt+\theta_{11})} \\
\tilde{r}_1 &= A^2 t_{12} t_{21} I_0 e^{i(wt+\theta_{12}+\theta_{21}+2\delta+\pi)} \\
\tilde{r}_2 &= A^2 r_{22} e^{i(\theta_{22}+2\delta+\pi)} \tilde{r}_1 = q\tilde{r}_1 \\
\tilde{r}_n &= q^{n-1} \tilde{r}_1
\end{aligned} \quad (1)$$

where $I_0$ is the magnitude of the incident wave, $r_{ii}, \theta_{ii}$ are the magnitude and phase coefficients for the reflected waves from $i^{th}$ layer back to $i^{th}$ layer, $t_{ij}, \theta_{ij}$ are the magnitude and phase coefficients for the transmitted waves from $i^{th}$ layer to $j^{th}$ layer, $A = e^{-\frac{2\pi}{\lambda}\kappa d}$, $\delta = nk_0 d$ are the propagation phase and loss caused by spacer layer with a thickness of $d$ and refractive index $\tilde{n} = n - i\kappa$. These coefficients at individual interfaces can be derived from numerical simulations by decoupling the model (Fig. 2(b)).[15] It should be noticed that since the material we chosen is a 0.54 Ω cm boron-doped silicon, whose permeability can be expressed by Drude model as:[16]

$$\varepsilon(\omega) = 11.7 - \frac{3.64 \times 10^{26}}{\omega^2 + 1.74 \times 10^{13} \omega i} \quad . \quad (2)$$

The high value of the imaginary part of the permeability can lead to a relative large $A$, hence the reflection waves of higher orders are dramatically declined, i.e., the magnitude of $\tilde{r}_2$ will be less than $0.01 I_0$, which makes that the overall reflection waves are mainly contributed by $r_0$ and $\tilde{r}_1$. From Eq. (1), we can obviously conclude that when $\theta_{12} + \theta_{21} + 2\delta + \pi - \theta_{22} = 2m\pi + \pi$ ($m \in Z$), the overall reflection will reach the minima at $f_1$=0.87 THz, $f_3$=1.54 THz and $f_5$=2.25 THz, since $r_0$ and $\tilde{r}_1$ will destructively interference with each other (black dash lines in Figs. 2(d) and 2(f)).

However, in order to achieve an ultra-broadband anti-reflection spectrum, the enhanced reflection caused by the constructive interferences between the destructive interferences should be tailored. Recent researches have revealed that the grating on the heavily-doped silicon can also reduce the reflection at certain frequencies.[12] The impedance-matching condition will be fulfilled, when the impedance of the grating is[11]

$$Z_{gra} = Z_0\sqrt{\cos\phi / n},$$

where $\phi = \sin^{-1}(m\lambda/(np))$ is the diffraction angle of the m$^{th}$ order diffraction and n is the refractive index of the silicon spacer. Through the two dimensional rigorous coupled wave analysis (2D-RCWA) method,[17] the diffraction efficiency of different orders is calculated, as demonstrated in Fig. 3(a). The gray zone in Fig. 3 indicted the frequency band where the [±1,0] order diffraction is dominant. The ideal impedance of the grating should be the combination of the [0,0] order impedance in the [0,0] order zone (red solid line in the white zone in Fig. 3(b)) and the [±1,0] order impedance in the [±1,0] order zone (blue solid line in the gray zone in Fig. 3(b)). We retrieve the effective impedance of the grating by adopting the S-parameter retrieval method[18] and optimize the parameters of the grating to make its impedance equal to the ideal impedance at the corresponding frequencies. After the optimization, the impedance-matching condition is thus reached at $f_2$=1.16 THz, $f_3$=1.54 THz and $f_4$=1.95 THz (highlighted as orange in Figs. 3(b) and 3(c)). So it can be concluded that the successive reflection dips of the absorber in a wide frequency region (Fig. 2(f)) is resulted from the impedance-matching condition of the grating superposed on the destructive interference of the multi-layer structure (black and red dash line in Figs. 2(d), 2(e) and 2(f)).

To verify the aforementioned analysis, the theoretical absorbance of designed absorber with two pairs of perfect electric walls and perfect magnetic walls bounding a propagation region was simulated via CST microwave studio®. As illustrated in Fig. 3 (a), five near-unity absorption peaks can be observed in the simulated absorption spectrum which has a good agreement with our analysis. Furthermore, the energy density distributions of the absorber at each absorption peak frequency are plotted in Figs.3 (b)-(f), which clearly reveal that the energy is concentrated at the grating/spacer interface at $f_2$, $f_3$, $f_4$ and the energy are captured in the spacer at $f_1$, $f_3$, $f_5$ due to the destructive interference. These distributions prove our analysis on how the destructive interference and the impedance-matching condition of the grating contributes to the absorption peaks.

## 3. FABRICATION AND MEASUREMENT

Standard wafer grinding, metal evaporation, optical lithography, and inductively coupled plasma (ICP) etching techniques were used to fabricate the multilayer absorber. First, a 500 μm-thick silicon wafer was grinded to 120 μm. A 20/180nm Ti/Au metallic plate was then evaporated onto the silicon wafer. On the other side of the wafer, the grating structure was finally formed by the optical lithography of the photo resist and the ICP etching.

We used the terahertz time domain spectroscopy (THz-TDS) system[19] to experimentally evaluate the behavior of the fabricated absorber by measuring the reflectance spectra. The THz-TDS system can produce terahertz waves in the range of 0.2-2.5 THz with an 8 GHz spectral resolution. The measured spectra in transverse electric (TE) and transverse magnetic (TM) mode (Fig. 5(a) and 5(b)) reveal five absorption peaks at the frequency 0.88 THz, 1.20 THz, 1.53 THz, 1.96 THz and 2.23 THz with absorbance of 99.77%, 95.82%, 99.98%, 99.55%, and 99.96%, respectively. Attributing to these five closely positioned absorption peaks, an ultra-wide frequency band is achieved from 0.75 THz to 2.41 THz, where the absorption is greater than 95%. Excellent agreements between simulation and experiment spectra are observed, with slight difference, which may be caused by the fabrication and measurement errors of the sample. The numerical simulations of absorption spectrum with different angles of incidence under TE and TM polarization state illustrated in Figs. 5(b) and 5(c) shows that the absorber can maintain an over 90% absorbance at a large incidence angle under both polarization states. Furthermore, owing to the symmetrical design of the device, the absorption response is insensitive to polarizations states of the normal incident wave.

## 4. CONCLUSION

In summary, we have theoretically purposed and experimentally verified a nearly perfect absorber in terahertz regime. The absorber is much simpler to fabricate and has a larger RAB than other broadband MPA in terahertz frequency. The fabricated absorber experimentally achieved a greater than 90% absorption over a frequency range of 1.8 THz. Destructive interference theory and impedance-matching condition of the grating are applied to elucidate the five successive absorption peaks at 0.88 THz, 1.20 THz, 1.53 THz, 1.96 THz and 2.23 THz. Owing to the symmetrical design, the absorber is also independent to the polarization states of the incident wave.

Furthermore, the absorber is also valid to a wide range of incident angles for both transverse electric (TE) and transverse magnetic (TM) polarizations.

## 5. ACKNOWLEDGEMENT

This work was partly supported by National Program on Key Basic Research Project of China (973 Program, 2014CB339806), Basic Research Key Project (12JC1407100), Major National Development Project of Scientific Instrument and Equipment (2011YQ150021) (2012YQ14000504), National Natural Science Foundation of China (11174207) (61138001) (61205094) (61307126), Shanghai Rising-Star Program (14QA1403100), Program of Shanghai Subject Chief Scientist (14XD1403000), and the Scientific Research Innovation Project of Shanghai Municipal Education Commission (14YZ093).

[a] Author to whom correspondence should be addressed. E-mail: xfzang@usst.edu.cn
[b] Author to whom correspondence should be addressed. E-mail: ymzhu@usst.edu.cn

**Figure Captions**

Figure 1 The structure of the sample:

(a) 3D Schematic diagram;

(b) SEM Photograph.

Figure 2 Analysis on the contribution of the destructive interference to the overall reflection:

(a) The interference model of the absorber;

(b) The simulated magnitudes of the reflection and transmission coefficients for the decoupled model.

(c) The simulated phases of the reflection and transmission coefficients for the decoupled model.

(d) The calculated phase difference between $r_0$ and $\tilde{r}_1$ from the decoupled model.

(e) The reflectance of the grating

(f) The reflectance of the whole structure

Figure 3 Analysis on the contribution of the grating to the overall reflection:

(a) The diffraction efficiencies for different orders calculated by 2D-RCWA method, gray zone indicates where the [1,0] order diffraction efficiency is greater than [0,0] order diffraction efficiency.

(b) The ideal and optimized impedance of the grating

(c) The reflectance of the grating, the anti-reflection of the grating caused by the impedance-matching condition is highlighted in orange.

Figure 4 The simulated results via CST microwave studio®

(a) the absorption spectrum;

(b)-(f) the energy density distributions in the *x-z plane* at y = 36.5μm at five absorption peaks; d

Figure 5

The comparisons of the experimental result and the simulated result in (a) TE mode (b) TM mode;

The dependences of the absorption spectra on incident angles in (a) TE mode (b) TM mode.

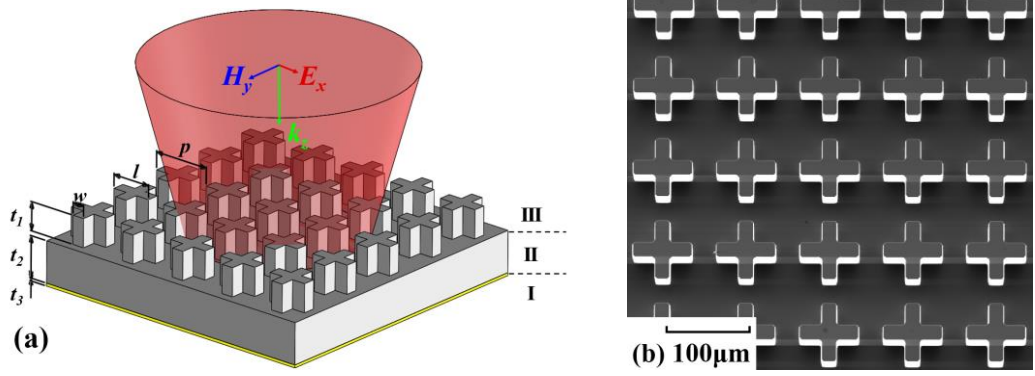

**Figure 1**

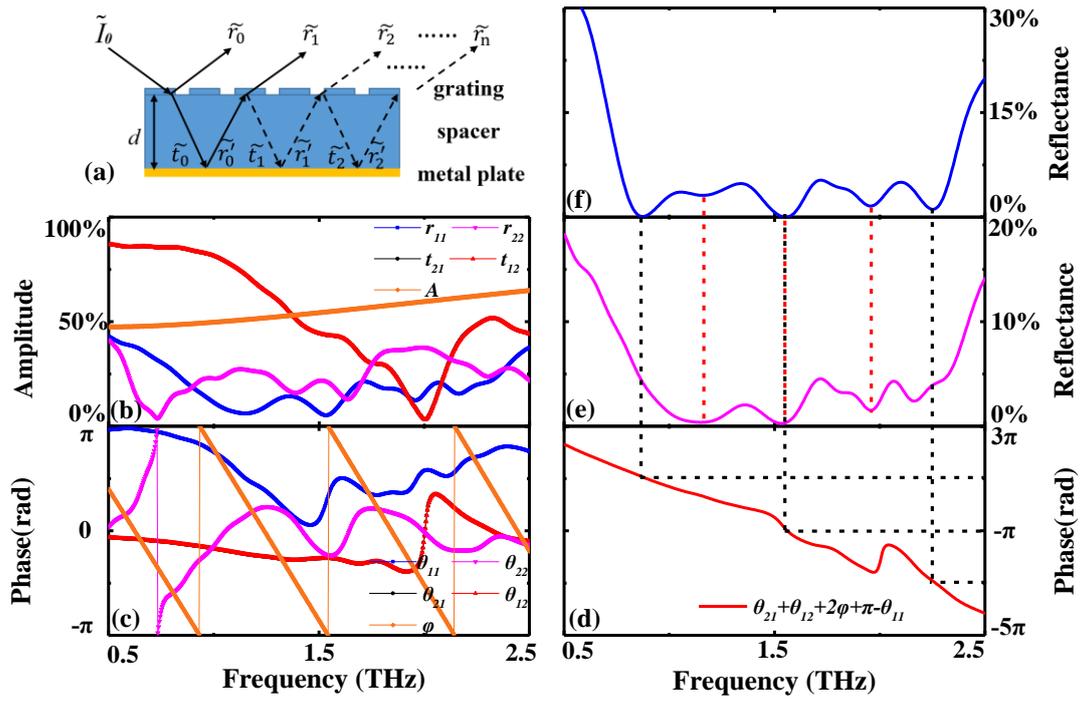

**Figure 2**

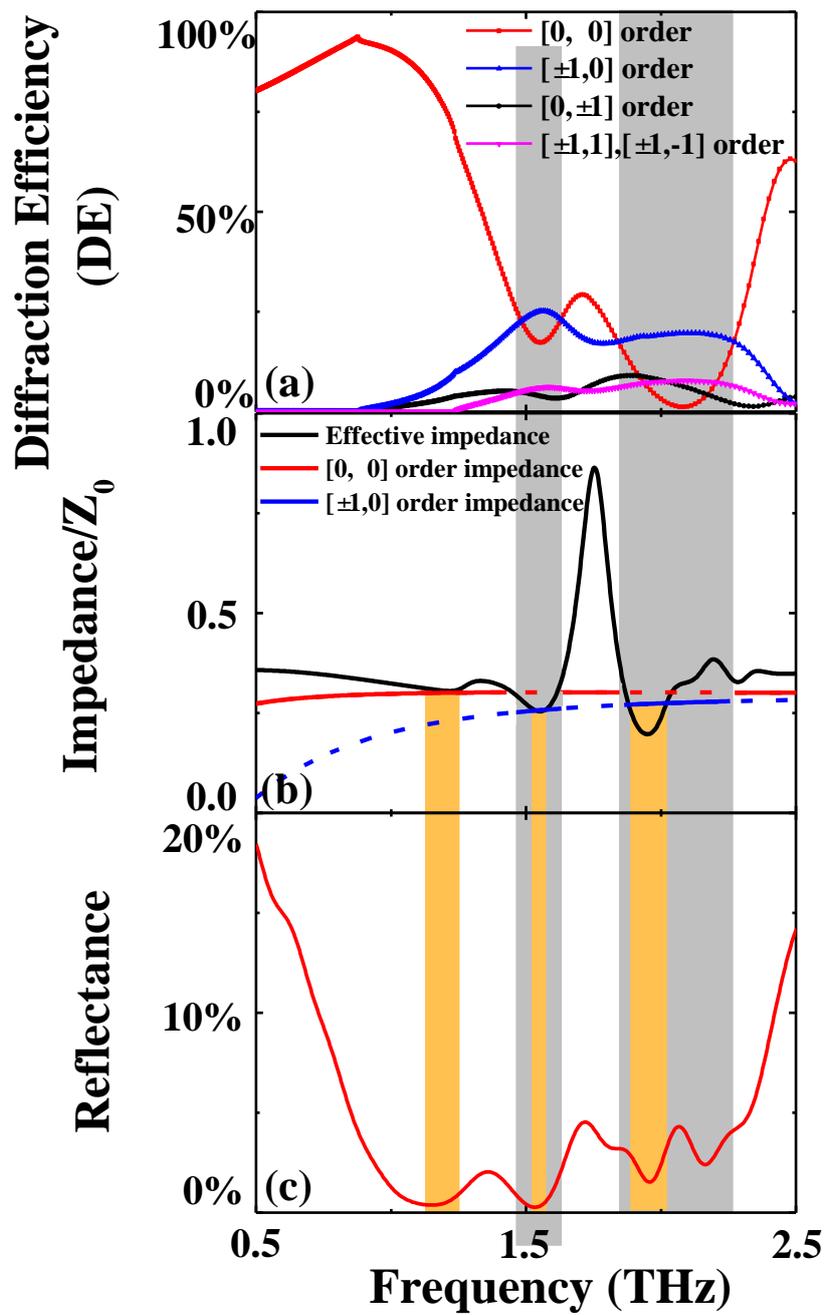

Figure 3

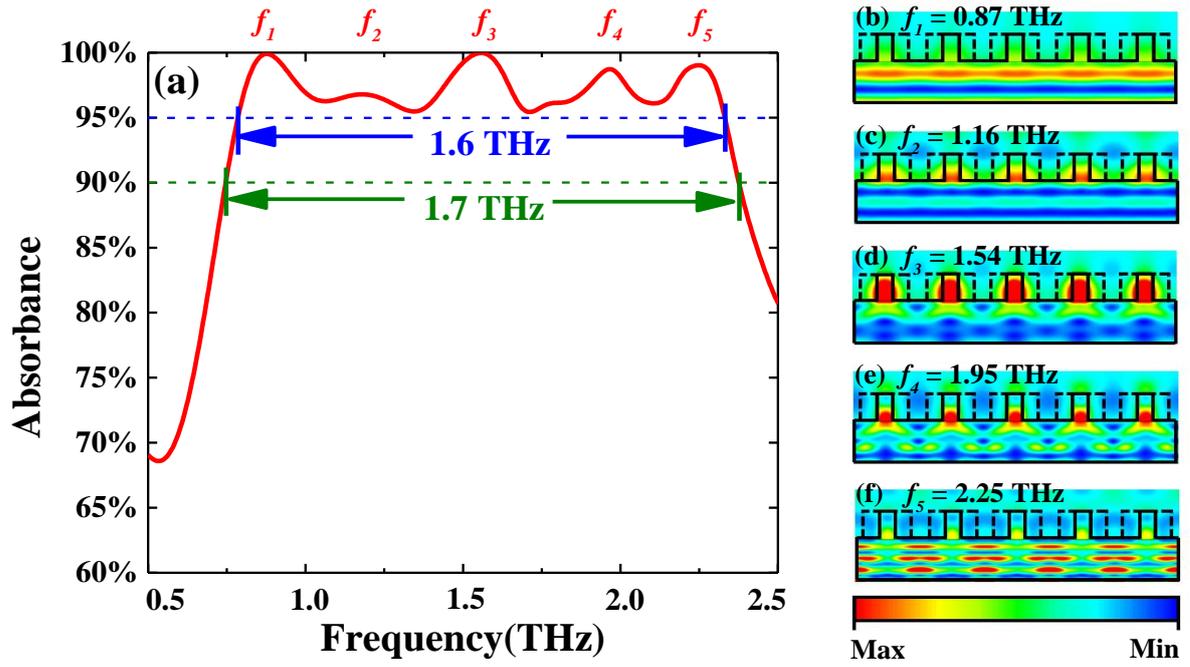

**Figure 4**

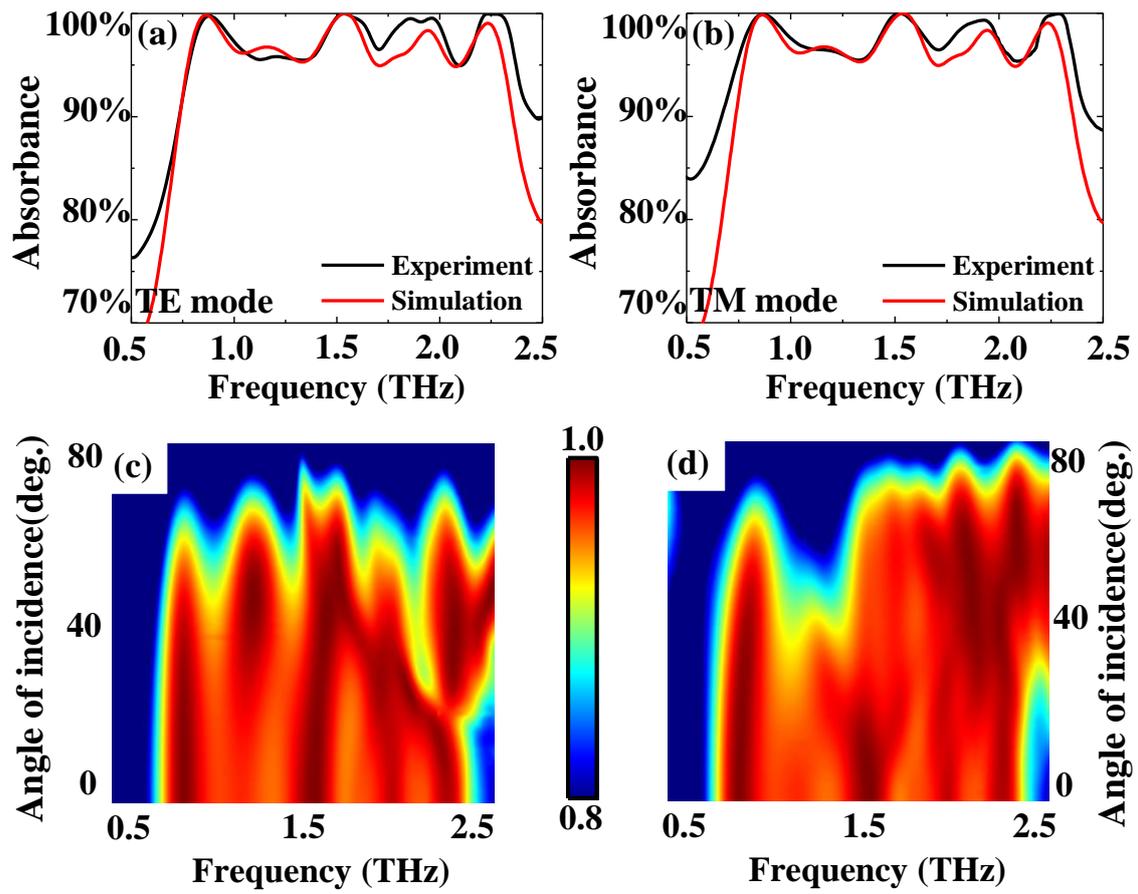

Figure 5